# Solutions on 1D and 2D Density Classification Problem Using Programmable Cellular Automata

Sudhakar Sahoo, Pabitra Pal Choudhury, Amita Pal,
Indian Statistical Institute, Kolkata, 700108, INDIA
Email: sudhakar.sahoo@gmail.com, pabitrapalchoudhury@gmail.com, pamita@isical.ac.in

Birendra Kumar Nayak,
P.G. Department of Mathematics,
Utkal University, Bhubaneswar-751004, INDIA
Email: bknatuu@yahoo.co.uk

*Abstract*—**This paper presents solutions to Density Classification Task (DCT) using a variant of Cellular Automata (CA) called Programmable Cellular Automata (PCA). The translation property as well as the density preserving property of fundamental CA rules in 1D and 2D, and the advantage of PCA are embedded together to obtain the DCT solution. The advantage of PCA over standard CA is reported. A general 2D DCT of arbitrary shapes and sizes, its applicability and its solution using PCA is newly introduced.**

## I. Introduction

Cellular Automata (CA) first introduced by J. v. Neumann (von Neumann, 1966) and further popularized by S. Wolfram by publishing a book entitled "A New Kind Of Science" (Wolfram, 2002) encourage other researchers to work in this area because of its regular, modular, simpler and easily cascadable computing structure. Further, the Parallelism property inherently embedded in the Cellular Automata tool dramatically increases the performance of any application that uses it.

CA can be considered as an abstract formalism for parallel computation. The computing process of CA is carried out by simple interconnected components. According to the standard definition (Wolfram, 2002) the components can be in a finite number of possible states, which are updated in parallel by the same updating rule. The updating rule depends only on the state of the component and on the state of its neighbors, thus no component can "observe" the whole system. It is usually easy to model biological and other complex systems with the help of CA.

On the other hand the problems which are intrinsically global in nature like the Density Classification Problem (also known as Density Classification Task (DCT)), first introduced by Packard, is a simple counting problem (Packard, 1988) but, is very difficult for any CA to get its solution, since it has neither the memory to keep a cumulative count nor the spatial range of perception to make a correct assessment. The Density Classification Problem is therefore attracted a great deal of interest to many CA researchers to test and measure the computing ability of CA.

Other motivations in solving the DCT has its root in solving other real-world problems such as image processing, analysis of Balanced Boolean functions, to test a matrix either dense or sparse, recognizing acid-base solution depending upon its PH values (Reynaga and Amthauer, 2003), finding two class pattern classifier which is further can be extended to solve k-class pattern classification problem (Maiti et al., 2006) etc.

All the work to date on this problem uses the natural approach of determining a rule's performance on using evolutionary algorithms (Mitchell et al., 1994, 1998; Crutchfield and Mitchell, 1995; Jullie and Pollack, 1998; Inverso et al., 2002), or a reverse engineering approach using the basin's of attraction (Bossomaier et al., 2000) or using analytic formulation of some specific CA rules (Land and Belew, 1995; Sipper et al., 1998; Maiti et al., 2006) to get an approximate solution. This paper on the other hand, presents a deterministic method to get an exact solution of both 1D as well as 2D DCT and doesn't use any heuristics as reported by other authors. Here, the authors propose a solution to the DCT by using a variant, called Programmable CA (PCA), in which the automaton is allowed to change the updating rule both in space and time, i.e. different components can use different updating rules and they are allowed to change their updating rule during the computation. The presented solutions arrive from a two-step procedure as a slight modification of the classical DCT definition originally introduced by Packard (Packard, 1998). More specifically, the initial CA configuration with the application of proper CA rules results in an intermediate CA configuration of 0's and 1's and then the PCA picks up the classification decision based on the pre-produced densities. Some examples of the presented 1D and 2D DCT algorithms and possible applications of the proposed PCA are also discussed.

The paper is organized as follows. Section II provides all the needed mathematical information on the basic concepts of both elementary and of Programmable Cellular Automata (PCA). Surveys on earlier work that elucidate DCT solution are given in section III. A formal definition of DCT and the Programmable Cellular Automata (PCA) rules those are used in the proposed algorithms are given in section IV. Section V and VI are about the proposed algorithms to solve 1D and 2D DCT problem respectively. Additionally section VI also defines other variety of 2D DCT, its practical application and proposes its solution using the concept of 2D PCA. Finally, section VII concludes the paper.

## II. BASIC CONCEPTS ON CA AND PCA

### A. Elementary CA both in 1D and 2D

A 1D standard Cellular Automaton (CA) (also called elementary CA) consists of two things: a row of "cells" and a set of "rules". The *row of cells* at any time-instant $t$ is represented by a vector $(x_1^t, x_2^t, ..., x_n^t)$ where $x_i^t$ denotes the bit in the $i^{th}$ cell $x_i$ at time-instant $t$. The bit in the $i^{th}$ cell at the "next" time-instant $(t+1)$ is given by a *local mapping* denoted by $f^i$ and all the cells change state at the same time. Local mapping means when the time comes for the cells to change state, each cell looks around and gathers information (READ operation) on its neighbors' states. Based on its own state Y (see fig-1 for 1D 3-neighborhood), its left neighbor's state X, its right neighbor's state Z and the rule of the CA, the cell decides (WRITE operation) what its new state should be. Thus the local rule is of the form,

$$x_i^{t+1} = f_i^t(X = x_{i-1}^t, Y = x_i^t, Z = x_{i+1}^t), i = 2,3,...,n-1$$

| X | Y | Z |

Figure 1. 1D neighborhood of radiuss 3

For the cells $x_1$ and $x_n$ it can be noted that different boundary conditions can be imposed such as Null, Periodic, Fixed, Adiabatic, Reflexive, Intermediate etc. In case of **Fixed boundary** the extreme cells are connected by a pre-assigned fixed logic 0/1 states. If the extreme cells are connected by logic 0 states then it is called **Null boundary** CA. In case of **Periodic boundary** the extreme cells are connected to each other and form a cycle. In case of **Adiabatic boundary** the missing neighbors states (or Virtual neighbors) are duplicate of the boundary cell values. In case of **Reflexive** boundary the value of left and right neighbors are same with respect to the boundary cell. In case of **Intermediate** boundary the value of the left (right) boundary will be the same as the cell value present at its next to next ((previous to previous) cell. Different 1D 3-neighborhood boundary conditions are shown below.

Fixed: $0(x_1, x_2, x_3, ..., x_{n-1}, x_n)0$ or $0(x_1, x_2, x_3, ..., x_{n-1}, x_n)1$

Or $1(x_1, x_2, x_3, ..., x_{n-1}, x_n)0$ or $1(x_1, x_2, x_3, ..., x_{n-1}, x_n)1$

Null: $0(x_1, x_2, x_3, ..., x_{n-1}, x_n)0$

Periodic: $x_n(x_1, x_2, x_3, ..., x_{n-1}, x_n)x_1$

Adiabatic: $x_1(x_1, x_2, x_3, ..., x_{n-1}, x_n)x_n$

Reflexive: $x_2(x_1, x_2, x_3, ..., x_{n-1}, x_n)x_{n-1}$

Intermediate: $x_3(x_1, x_2, x_3, ..., x_{n-2}, x_{n-1}, x_n)x_{n-2}$

If the same CA rule determines the "next" bit in each cell of a CA, the CA will be called a **Uniform Cellular Automaton**; otherwise it will be called a **Hybrid Cellular Automaton**. In case of 1D, 3-neighborhood, 2-state the number of all possible uniform CA rules is $2^8 = 2^{2^3} = 256$. These rules can be enumerated using Wolfram's naming convention (Wolfram, 1986) from rule number 0 to rule number 255 and can be represented by a 3-variable Boolean function. The algebraic expression (Wegener, 1987) for any 3-variable Boolean function contains term including X, Y, Z in general. Under this perspective these three rules, which will be used in latter sections, are considered fundamental and are shown in fig-2. The CA evolution (also known as Space-time diagram) of Rule Z (equivalently Rule 170) starting from an initial configuration is shown in fig-3. It may be noted that applying Rule Z clustering of 1's in right side have been shifted to left side.

| $XYZ$: | 111 | 110 | 101 | 100 | 011 | 010 | 001 | 000 |
|---|---|---|---|---|---|---|---|---|
| Rule X or Rule 240: | 1 | 1 | 1 | 1 | 0 | 0 | 0 | 0 |
| Rule Y or Rule 204: | 1 | 1 | 0 | 0 | 1 | 1 | 0 | 0 |
| Rule Z or Rule 170: | 1 | 0 | 1 | 0 | 1 | 0 | 1 | 0 |

Figure 2. 3-fundamental 1D CA rules with 3-neighborhood structure

```
Initial   :  0 0 0 0 0 0 1 1 1 1 1
time = 1  :  0 0 0 0 0 1 0 1 1 1 1 1
time = 2  :  0 0 0 0 0 1 0 1 0 1 1 1
time = 3  :  0 0 0 0 1 0 1 0 1 0 1 1
time = 4  :  0 0 0 1 0 1 0 1 0 1 0 1
time = 5  :  0 0 1 0 1 0 1 0 1 0 1 0
time = 6  :  0 1 0 1 0 1 0 1 0 1 0 0
time = 7  :  1 0 1 0 1 0 1 0 1 0 0 0
time = 8  :  1 1 0 1 0 1 0 1 0 0 0 0
time = 9  :  1 1 1 0 1 0 1 0 0 0 0 0
time = 10 :  1 1 1 1 0 1 0 0 0 0 0 0
time = 11 :  1 1 1 1 1 0 0 0 0 0 0 0
```

Figure 3. Shows the Space-time diagram of Rule Z (or Rule 170)

In 2D nine-neighborhood elementary CA: the next state of a particular cell is obtained by the current state of itself and eight cells in its nearest neighborhood (also referred as Moore neighborhood) as shown in fig-4.

| 64 | 128 | 256 |
|---|---|---|
| 32 | 1 | 2 |
| 16 | 8 | 4 |

Figure 4. Shows a naming scheme for 2D nine- neighborhood fundamental CA rules

Such dependencies are accounted by various rules or transition functions. For the sake of simplicity, in this section we take into consideration only nine fundamental rules. A specific rule convention that is adopted in our previous paper (Choudhury et al., 2008) is shown in fig-4 where, the central box represents the current cell (i.e. the cell being considered) and all other boxes represent the eight nearest neighbors of that cell. The number within each box represents the rule number characterizing the dependency of the current cell on that particular neighbor only. Rule 1 characterizes dependency of the central cell on itself alone whereas such dependency only on its top neighbor is characterized by rule 128, and so on. Similar to 1D case, these nine rules are named as fundamental rules. As reported earlier in our previous paper, (Choudhury et al., 2008) these rules are 1, 2, 4, 8, 16, 32, 64, 128, and 256. The application of these rules can be realized on a problem matrix or information matrix where every entry is either 0 or 1. It may be mentioned that instead of applying the same rule to each entry of the problem matrix, it is admissible to apply different fundamental rules to different entries at the same time. While the former characterizes the **uniform CA** the latter characterizes **hybrid CA** in 2D. Also the different boundary conditions such as Null, Periodic, Fixed, Adiabatic, and Reflexive etc defined in 1D CA can be defined in a similar way in case of 2D.

**Illustration (For Uniform CA)**

In this example, Rule 128 is applied uniformly to each cell of a problem matrix of order (3 x 4) with periodic boundary condition. It may be noted that on applying Rule 128 the rows of the problem matrix being shifted from top to bottom in a periodic way as shown in fig-5.

$$\begin{pmatrix} 0 & 0 & 1 & 0 \\ 1 & 1 & 1 & 0 \\ 1 & 0 & 1 & 1 \end{pmatrix}_{(3\times 4)} \xrightarrow{Rule\ 128\ (Periodic\ Boundary)} \begin{pmatrix} 1 & 0 & 1 & 1 \\ 0 & 0 & 1 & 0 \\ 1 & 1 & 1 & 0 \end{pmatrix}_{(3\times 4)}$$

Figure 5. Shows Rule 128 is applied to all cells that is each cell will change its state by the states of its top neighboring cells using the periodic boundary definition.

**Illustration (For hybrid CA)**

Here is an example of a hybrid CA in null boundary condition where 3 rules (Rule 2, Rule 8, and Rule 16) are applied in 3 different rows (1$^{st}$, 2$^{nd}$ and 3$^{rd}$ rows respectively) in the problem matrix.

$$\begin{pmatrix} 0 & 0 & 1 & 0 \\ 1 & 1 & 1 & 0 \\ 1 & 0 & 1 & 1 \end{pmatrix}_{(3\times 4)} \xrightarrow{Rule\ 2, Rule\ 8, Rule\ 16\ (Null\ Boundary)} \begin{pmatrix} 0 & 1 & 0 & 0 \\ 1 & 0 & 1 & 1 \\ 0 & 0 & 0 & 0 \end{pmatrix}_{(3\times 4)}$$

Figure 6. Shows 3 rules (Rule 2, Rule 8, and Rule16) are applied in 3 different rows (1$^{st}$, 2$^{nd}$ and 3$^{rd}$ rows respectively) in the problem matrix.

*B. Programmable CA (PCA)*

On using software if we try to implement a local CA rule, both READ and WRITE operations are essential but one can also capture the CA evolutions by some specific hardware architecture as shown in fig-7. This architecture involves single bit READ or WRITE operations automatically and does not involve memory in usual sense, unless we want to store the successive CA evolutions (or known as Space–time diagram) in some memory space. For this type of architecture a single clock cycle is required for a vector $(x_1, x_2, x_3, ..., x_{n-1}, x_n)$ (i.e. an n-bit CA) to get its next global state.

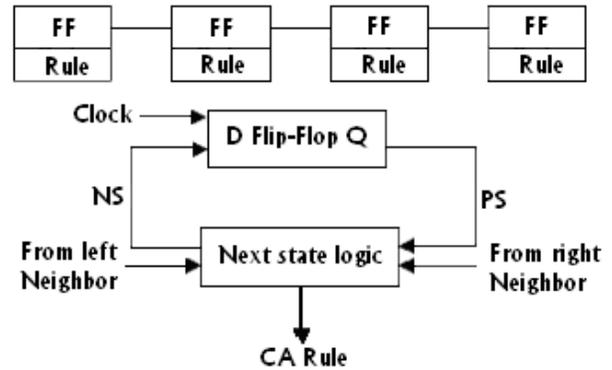

Figure 7. Shows a 4-cell CA and the way each cell change its state using 3 neighborhoods CA rule

Both hardware and software implementations are highly essential because using dedicated hardware architecture is expensive and not commonly available in the market. So for the analysis of CA evolutions researchers normally simulate it on a standard PC with the help of some program.

As reported in (Nandi et al., 1994) PCA allows spatial and temporal variations in the state transition rules within a CA, according to some external control scheme. This can be achieved by dynamically changing the CA rules. Through an appropriate selection of CA rules and the trio of logic gate/connection/control signal wirings, a number of rules can be programmed into the operation of the PCA. Thus the PCA architecture is very much flexible in emulating different hybrid CA configurations. Similar to a conventional PCA, **Self Programmable CA (SPCA)** has a localized state transition neighborhood for updating cell states and a new rule selection neighborhood for selecting rules (Guan and Tan, 2004). The rules are programmable by the control lines. In this paper the word PCA is used for both Programmable Cellular Automata and Self Programmable Cellular Automata depending on their context.

PCA has practical engineering applications and in this paper the concept of PCA is used to solve both 1D and 2D DCT problem. Since fig-7 demonstrates the natural CA architecture, one can reduce considerably the so-called READ and WRITE operations using PCA.

## III. SURVEY ON DENSITY CLASSIFICATION TASK (DCT)

Density Classification Task as reported by (Packard, 1988) is a simple computing problem in which the CA starts with an initial configuration and forms a configuration with either all 1's or all 0's depending on which had the higher initial density. For example, if the CA's initial state contains 60% 1's and 40% 0's then there are more 1's than 0's so the CA should become 100% 1's and remain in that state. Like wise, if the initial state contains 30% 1's and 70% 0's then the CA should become all 0's.

It has been proved that no 1D two-state CA can be constructed, which classifies binary strings according to their densities of 1's and 0's (Land and Belew, 1995). Further it has been demonstrated (Capcarrere et al., 1996) that a solution to the density classification problem does exist, defining a different output in comparison to that of Packard and (Capcarrere and Sipper, 2001) demonstrated that a rule, which resolves the density classification problem, for a one-dimensional elementary CA, has to satisfy two conditions:

1. The density of the initial configuration must be preserved over time.

2. The rules table must exhibit density of 0.5 implying that it should be a balanced rule.

A group of researchers of the Santa Fe institute uses Genetic Algorithm (GA) to evolve CA rules, which are able to solve the problem of the Density Classification. Results of Selected Rules can be found in (Mitchell et al., 1994, 1998; Crutchfield and Mitchell, 1995; Jullie and Pollack, 1998; Bossomaier et al., 1999).

An alternative, however, is the use of the algorithm developed by (Wuensche and Lesser, 1992; Wuensche, 1994) and (Chaudhuri et al., 1997; Maji and Chaudhuri, 2008) for running a CA rule backwards. In other words, find a CA rule such that the whole of configuration-space is divided into precisely two basins of attraction; one of which consists of all rules with density greater than the critical density $\rho_c$ and has as its attractor the state of all 1's, and the other consisting of all other configurations and having as its attractor the state of all 0's.

This then raises the possibility that one can now avoid some of the difficulties, both conceptual and computational, of the standard 'forwards' approach. For example, one might be able to more efficiently estimate the performance of a rule not by picking CA configurations at random and following each of them forward to their attractor. The main problem in this approach is the wide variation in quantitative results obtained when sampling different parts of the attractor basin.

In (Gabriele, 2005) a genetic algorithm evolves 1D CA in order to perform the classical main task is reported. First time this paper uses higher number of states called multi-states CA and solves the density classification problem. Their result shows that there is a substantial homogeneity between elementary CA and multi state CA.

In (Fuks, 1997) a pair of elementary rules, namely the "traffic rule" 184 and the "majority rule" 232, performs the density classification task perfectly. The main result of their paper is the following proposition.

**Proposition 1** Let s be a configuration of length L and density $\rho$, and let n=$\lfloor (L-2)/2 \rfloor$, m=$\lfloor (L-1)/2 \rfloor$. Then $F_{232}^m(F_{184}^n(s))$ consists of only 0's if $\rho < 1/2$ and of only 1's if $\rho > 1/2$. If $\rho = 1/2$, $F_{232}^m(F_{184}^n(s))$ is an alternating sequence of 0 and 1, that is 01010101.

Although the solution proposed by (Fuks, 1997) performs the task in L time steps, it is straightforward to construct a faster algorithm, providing that we allow rules of large radius. If $f$ is a radius-1 rule, then $f^n$, the rule iterated $n$ times, is itself a CA rule of radius $n$. Therefore, the pair $(g,h)$, where $g = f_{184}^n$ and $h = f_{232}^n$, performs the classification task in L/n time steps, assuming that we iterate both $g$ and $h$ for L/2n time steps.

On utilizing the characteristic of CA rather than any statistical approach involves GA, Maiti et al. have given an analytical framework for 1D DCT solution (Maiti et al., 2006). In their work Best Rule Vector (BRV) has been derived from the analysis of Rule Vector Graph (RVG) generated from the Rule Vector (RV) of a CA. Also they have analyzed the error, obtained in the DCT solution by introducing two types of error primary and secondary, then tried to reduce the error by extending the neighborhood to k (k=3, 5, 7, 9) in 1D CA.

A solution of 2D DCT can be found in (Reynaga and Amthauer, 2003). Result of this paper shows that, for two-dimensional task, a perfect rule does not exist. However, their experiments show a good performance even when radius of neighborhood is minimum.

Another interesting question is to design a general algorithm to solve 1D and 2D density classification problem according to an arbitrary critical density $\rho_c$. To the best of our knowledge no earlier work exists that discusses this problem. Thus apart from solving the conventional DCT additionally our efforts in this paper would be to design and develop the general DCT problem and provide its solution with the help of PCA.

## IV. A NEW DCT DEFINITION AND PROPERTIES OF 1D AND 2D FUNDAMENTAL CA RULES

The presented solutions in this paper both for 1D and 2D arrive from a two-phase procedure as a slight modification of the classical DCT definition originally introduced by Packard (1988). For our purpose an equivalent density classification problem can be defined as follows:

**Definition:** The initial CA configuration with the application of proper CA rules results in an intermediate CA configuration of 0's and 1's and then the PCA picks up the classification decision based on the pre-produced densities.

This new way of defining density classification problem is equivalent to the original problem defined by Packard (Packard, 1988) as given earlier in section III. This is because of the existence of two trivial CA rules; Rule 0 and Rule $(2^{2^k} -1)$ for a neighborhood of radius $k$. For example when $k=3$ the Rule number is 255 in case of elementary 1D CA and in case of 2D $k=9$ so rule number is $(2^{512}-1)$. An interesting property of these trivial rules says that: Starting from any random configuration, Rule 0 and Rule $(2^{2^k}-1)$ each directly evolves to a quiescent configuration of cell values all 0's or all 1's respectively.

### A. Translation of images and density preservation by 1D fundamental rules

In table I, two fundamental 1D CA rules, Rule X in Algebraic Normal Form (ANF) (Wegener, 1987) equivalently Rule 240 in Wolfram convention (Wolfram, 1986) and Rule Z (equivalently Rule 170) has the ability to translate 1's from left to right and from right to left respectively. Again the identity rule, Rule Y (ANF) equivalently Rule 204 (Wolfram convention) can preserves the number of 1's and 0's constant throughout the evolution. Thus the combined effect of these three rules can serve both the properties: translation as well as density preservation as shown in table I.

TABLE I. DIRECTIONS OF TRANSLATION OF AN IMAGE ON APPLYING 3-FUNDAMENTAL 1D CA RULE

| Direction of the translation of the image | Rules to be applied |
|---|---|
| Left ← | Rule X (ANF) or Rule 240 (Wolfram) |
| No translation but densities of 0's and 1's are preserved | Rule Y (ANF) or Rule 204 (Wolfram) |
| Right → | Rule Z (ANF) or Rule 170 (Wolfram) |

### B. Translation of images and density preservation by 2D fundamental rules

It is reported in (Choudhury et al., 2008) that the fundamental rules of fig-4 namely rule 2, 4, 8, 16 and their transpose rules 32, 64, 128, 256 respectively, cause translation of the image in the directions mentioned against each rule in the table-II. The fundamental rule: Rule 1 being an identity rule satisfies density preservation (number of 1's and 0's constant throughout the evolution) property.

For applying 2D CA fundamental rules for translation of images, we take a binary matrix of size $(100\times100)$. We map each element of the matrix to a unique pixel on the screen (Using BGI graphics of Turbo C++ 3.0) and we color a pixel White for 0 and Black for 1 for the matrix elements. We apply 2D CA fundamental rules uniformly on that $(100\times100)$ matrix and each time the rule is applied using the changed matrix and a new image is redrawn. Initial image is taken by making some elements of the $(100\times100)$ matrix as 1, in a suitable fashion. We call this initial image as 'seed image'. The extent of shift is dependent on the number of repetitions of the application of the fundamental rules as illustrated in the fig-8 followed by table-II.

TABLE II. DIRECTIONS OF TRANSLATION OF AN IMAGE ON APPLYING 9-FUNDAMENTAL 2D CA RULE

| Direction of the translation of the image | Rules to be applied |
|---|---|
| Top (↑) | Rule 8 |
| Bottom (↓) | Rule 128 |
| Left (←) | Rule 2 |
| Right (→) | Rule 32 |
| No translation but densities of 0's and 1's are preserved | Rule 1 |
| Top-Left (↖) | Rule 4 |
| Top-Right (↗) | Rule 16 |
| Bottom-Left (↙) | Rule 256 |
| Bottom-Right (↘) | Rule 64 |

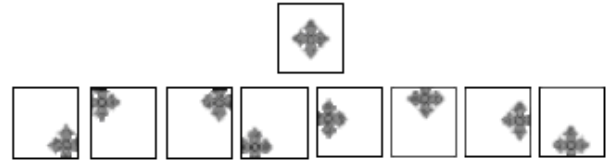

Figure 8. Shows the translation of an image after application of CA rules 64, 4, 16, 256, 2, 8, 32 and 128 to a given image

## V. SOLUTION OF 1D DCT USING PROGRAMMABLE CELLULAR AUTOMATA

### A. Algorithm to solve 1D Density Classification Problem

The algorithm discussed here to solve 1D DCT problem can be divided into two-phases, a preprocessing phase and a decision phase. In the preprocessing phase the initial CA configuration of length $n$ is evolved by CA rules to reach at a particular type of configuration after which a decision regarding the densities can be made in the decision phase. Just like Sieve of Eratosthenes technique to filter prime numbers and composite numbers the preprocessing phase in the proposed algorithm separates 1's and 0's to form separate clusters. One such intermediate configuration to be used in the pre-processing phase of the proposed 1D DCT algorithm is shown in fig-9.

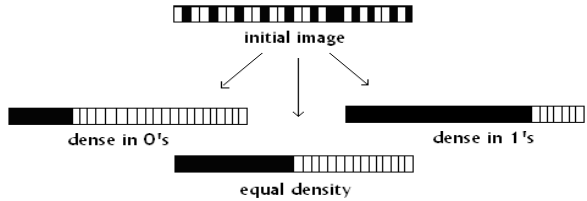

Figure 9. Possible intermediate outputs for taking decision in 1D DCT algorithm

Fig-9 shows that starting from an initial 1D configuration $(x_1, x_2, x_3, ..., x_{n-1}, x_n)$ of length $n$, we need some CA rules, which translates all 1's towards left, and all 0's towards right with a condition that number of 0's and 1's should not be changed throughout the evolution process. To achieve translation, 2 fundamental CA rules Rule X and Rule Z and for density preservation Rule Y of table II, can be used because they satisfies these requirements. Also some special care should be taken for different boundary conditions for example, in Periodic boundary the 1's those translates to wards left they should not again return back from the right end. Further, it may be mentioned here that the effect of an arbitrary rule vector can be captured by the combined effect of 3 fundamental rules X, Y or Z or their complements. This has been exemplified in fig-10 stating that at different neighborhood situation, rules X, Y or Z or their complements are invoked and as a result, rule 226 is obtained. Again selection of these rules X, Y or Z or their complements for a particular neighborhood situation may not be unique. For example in fig-10 the "XYZ" values $"110" \rightarrow "1"$ can be obtained by applying any one of the rule say either Rule X or Y or $Z^c$ (complement of Z).

Table III shows the list of 1D CA rules those can be used in the preprocessing phase proposed in the 1D DCT algorithm. These rules, which achieve both translation and density preservation, as discussed above can be obtained by the following logic:

For all the cells from $x_2$ to $x_{n-1}$ except the boundary the output requirements for all possible 3-neighborhood situations are given in fig-10. As an example for the neighborhood situation 111, 110, 100, and 000 where, 1's if any are present in the left side and 0's if any are in the right side so there is no need to change the state for this type of neighborhoods. But for other neighborhood situation 101, 011, 010, and 001 the 1's are followed by 0's so there is a need for translation of 1's from right to left. The complete change of all these eight situations is shown in fig-10 from which the desired rule vector can be constructed.

| XYZ: | 111 | 110 | 101 | 100 | 011 | 010 | 001 | 000 |
|---|---|---|---|---|---|---|---|---|
| Requirement: | 111 | 110 | 110 | 100 | 101 | 100 | 010 | 000 |
| Rule 226: | 1 | 1 | 1 | 0 | 0 | 0 | 1 | 0 |
| Fundamental rules | X/Y/Z | X/Y/$Z^c$ | X/$Y^c$/Z | $X^c$/Y/Z | X/$Y^c$/$Z^c$ | $X/Y^c$/Z | $X^c$/$Y^c$/Z | X/Y/Z |

Figure 10. Shows the CA rule 226 obtained by considering Translation as well as Density Preservation property of 3 fundamental rules X, Y or Z

From fig-10 collecting the middle cell values from the requirement row the 1D CA rule that achieves both translation as well as density preservation is the rule vector "11100010", which is rule 226 (by Wolframs Naming Convention).

TABLE III. RULES USED TO SOLVE 1D DCT PROBLEM

| (1) Cells | (2) If the left neighbour of $x_1 = 0$ then apply any rule to $x_1$ | (3) If the left neighbour of $x_1 = 1$ then apply any rule to $x_1$ | (4) Rules for all other cells $x_2$ to $x_{n-1}$ | (5) If the right neighbour of $x_n = 0$ then apply any rule to $x_n$ | (6) If the right neighbour of $x_n = 1$ then apply any rule to $x_n$ |
|---|---|---|---|---|---|
| Rules | 14, 30, 46, 62, 78, 94, 110, 126, 142, 158, 174, 190, 206, 222, 238, 254 | 224, 225, 226, 227, 228, 229, 230, 231, 232, 233, 234, 235, 236, 237, 238, 239 | 226 | 64, 66, 72, 74, 96, 98, 104, 106, 192, 194, 200, 202, 224, 226, 232, 234 | 128, 129, 132, 133, 144, 145, 148, 149, 192, 193, 196, 197, 208, 209, 212, 213 |

The logic to obtain the 1D CA rules for both left and right boundary cells of the vector $(x_1, x_2, x_3, ..., x_{n-1}, x_n)$ are as follows. If the value of left neighbor of the first cell $x_1$ is 0 then out of eight possible 3-neighborhood situations only four situations 011, 010, 001, and 000 where first bit is "0" will be valid for $x_1$ as shown in fig-11. Out of this four, as the value of the first cell $x_1 =1$ for the neighborhood situation 011 and 010 thus for density preservation there is no need to translate this 1 to wards left. So for these neighborhoods 011 and 010 we need a CA rule such that after applying the rule the requirements should be again 011 and 010 respectively. But for the string 001 the right side 1 can move to the cell position $x_1$. So for this neighborhood after applying the rule the requirement should be 010. Similarly after applying the CA rule to the neighborhood 000 the requirement should obviously be 000. Thus from the requirement row given in fig-11, the CA rule can be computed by assigning the values 1, 1, 1 and 0 respectively corresponding to these four neighborhoods and the other four neighborhood situations 111, 110, 101, and 100 where the left neighbor of $x_1$ is 1 can be assigned by 0 or 1. In this way 16 possible CA rules can be obtained and some of these rule vectors are 00001110=14,

00011110=30 and so on. The complete list of these rules is given in column (2) of table III.

*Rules if left neighbour of* $x_1 = 0$

$0x_1x_2$ :        111  110  101  100  011  010  001  000
Re *quirement* :    –    –    –    –    011  010  010  000
16 *Rules* :       0/1  0/1  0/1  0/1   1    1    1    0

Figure 11. Show the CA rules those can be used for the boundary cell $x_1$ if its left neighbour value is 0.

Fig-12 shows 16 other rules that can be obtained by fixing the left neighbor of $x_1 = 1$. The complete list of these rules in Wolfram's naming convention is given in column (3) of table III.

*Rules if left neighbour of* $x_1 = 1$

$1x_1x_2$ :        111  110  101  100  011  010  001  000
Re *quirement* :   111  110  110  100   –    –    –    –
16 *Rules* :        1    1    1    0   0/1  0/1  0/1  0/1

Figure 12. Show the CA rules those can be used for the boundary cell $x_1$ if its left neighbour value is 1.

Similarly all other rules applicable for the boundary cell $x_n$ can be found out by considering its right neighbour value either 0 or 1 as shown in fig-13 and the complete list of these rules are given in column (5) and (6) of table III.

*Rules if right neighbour of* $x_n = 0$

$x_{n-1}x_n0$ :    111  110  101  100  011  010  001  000
Re *quirement* :    –   110   –   100   –   100   –   000
16 *Rules* :       0/1   1   0/1   0    0/1   0   0/1   0

(a)

*Rules if right neighbour of* $x_n = 1$

$x_{n-1}x_n1$ :    111  110  101  100  011  010  001  000
Re *quirement* :   111   –   101   –   101   –   001   –
16 *Rules* :        1   0/1   0   0/1   0   0/1   0   0/1

(b)

Figure 13. Shows the CA rules used for boundary cell $x_n$. (a) gives the rules if the right neighbor of $x_n$ =0 and (b) computes the rules if it is 1.

Thus the main result of the above discussion gives a set of rules those are listed in table III that can be used in the pre-processing phase of the 1D DCT algorithm to reach at a configuration like fig-9.

After preprocessing phase is over depending on the length of CA is either even or odd a decision is made in the decision phase. For an odd length CA only two outputs are possible, which is either dense in 0's or dense in 1's. In this case only looking the value of the middle cell $x_{(n+1)/2}$ is either 0 or 1, it is possible to take the decision of classification. In case of an even length CA, looking two cells value $x_{n/2}$ and $x_{(n/2)+1}$ situated at the middle and the (middle+1) positions respectively one can take the decision for their densities. It may be mentioned here that in the decision phase two types of outputs are possible for DCT. In first type, one can print the output as the starting CA is dense in 0's or it is dense in 1's for odd length CA in addition with another possibility to print equal density in case of even length CA. This way of output is required for the DCT problem defined by (Capcarrere et al., 1996). In the second type, knowing the information that the CA is either dense in 0's or dense in 1's a trivial rule: Rule 0 or Rule 255 may be applied to get an output of 1D DCT defined by (Packard, 1988). Thus in this way it can able to solve both the original DCT defined by (Packard, 1988) as well as the DCT defined by (Capcarrere et al., 1996).

The proposed algorithm is very much common and compatible in order to handle different boundary conditions such as null, periodic, fixed, adiabatic etc. For example in case of null boundary condition where the extreme cells are connected to logic-0 states only the CA rules given in Column (2), (4) and (5) of table III will be used. Similarly for periodic boundary condition (where the extreme cells are connected to each other) depending upon the neighboring values of $x_1$ and $x_n$ different rules from table III can be selected. In addition to solve the 1D DCT problem the proposed 1D algorithm can act as a basis that can be used to develop the 2D DCT algorithm.

The proposed two-phase algorithm to solve 1D DCT problem using the CA rules from table III is as follows:

### 1) *Two-Phase Algorithm to solve 1D DCT*

/\* Pr *eprocessing Phase* : Re *aching to an intermediate configuration like fig* −9 \*/

1. Input: An initial 1D CA configuration $(x_1, x_2, x_3, ..., x_{n-1}, x_n)$ of length *n*.
2. $i = 1$
3. *while* (i < n)
4. {
/\* *Rules for Boundary cells in different boundary conditions as shown in table III* \*/
// *Rules for the first cell* $x_1$
5.   *if* the left *neighbour* of $x_1 = 0$ *then*
6.     *Apply* any rule to $x_1$ out of 16 possible rules given in column (2) of table III
7.   *else if* the left *neighbour* of $x_1 = 1$ *then*
8.     *Apply* any rule to $x_1$ out of 16 possible rules given in column (3)
// *Rules for the last cell* $x_n$
9.   *if* (the right *neighbour* of $x_n = 0$) *then*
10.    *Apply* any rule to $x_n$ out of 16 possible rules given in column (5)
11.  *else if* (the right *neighbour* of $x_n = 1$) *then*
12.    *Apply* any rule to $x_n$ out of 16 possible rules given in column (6)
/\* *Rule for all other cells from* $x_2$ *to* $x_{n-1}$ \*/
13.  *Apply* Rule 226 (*see* column (4)) uniformly to all the cells from $x_2$ to $x_{n-1}$
14. *if* (no updation taking place in all the cells) *then exit* from the *while* loop
15. *else* (*i*=(*i*+1)) and *continue*
16. }

/* Decision phase : For taking a decision of density */

17. if n is odd
18. {
19.   if (the bit at position $(n+1)/2 = 1$) then
20.     Apply Rule 255 uniformly to all the cells
21.     print "the CA is dense in 1's".
22.   else if (the bit at position $(n+1)/2 = 0$) then
23.     Apply Rule 0 uniformly to all the cells
24.     print "the CA is dense in 0's".
25. }
26. else if n is even
27. {
28.   if (two consecutive bits at positions $(n/2)$ and $(n/2)+1 = 00$) then
29.     Apply Rule 0 uniformly to all the cells
30.     print "the CA is dense in 0's".
31.   else if (two consecutive bits at positions $(n/2)$ and $(n/2)+1 = 10$) then
32.     print "equal density in 0's and 1's"
33.   else if (two consecutive bits at positions $(n/2)$ and $(n/2)+1 = 11$) then
34.     Apply Rule 255 uniformly to all the cells
35.     print "the CA is dense in 1's".
36. }

### 2) Explanation of the algorithm

Step 1 to Step 16 of the algorithm is the preprocessing phase and generates the required intermediate image like fig-9. Starting from the initial configuration of length $n$, in the worst case CA is evolved up to $(n-1)$ times by step 3. In each evolution process, Rule 226 is applied to all the cells except the boundary and to the boundary cells ($x_1$ and $x_n$), depending on the neighborhood values rules from table III can be selected as shown in steps 5 to 8.

For an odd length CA only two outputs are possible (Steps 17 to 25), which is either dense in 0's or dense in 1's. In this case only reading the middle cell value of the vector $(x_1, x_2, x_3, ..., x_{n-1}, x_n)$ one can take the decision of their densities.

For an even length CA, the decision for density classification is made on reading two cell values situated at the middle and the (middle+1) positions of the vector $(x_1, x_2, x_3, ..., x_{n-1}, x_n)$ (Steps 26 to 36).

It can be proved easily that, after the preprocessing phase of the 1D DCT algorithm, the string of two consecutive bits in the middle and (middle+1) position of the vector $(x_1, x_2, x_3, ..., x_{n-1}, x_n)$ can never be "01". If so then depending on the cell value situated at (middle-1) position of the vector either 0 or 1, the (middle) cell whose value is now "0" and its right neighbor (middle+1) value is "1" the neighborhood situation is either 001 or 101. Thus by step 13 when Rule 226 (see fig-13) is applied to the (middle) cell it will update its own state from 0 to 1. Thus by steps 14 and 15 of the algorithm at least one more evolution takes place, which tells that, the preprocessing phase is not yet over and contradicts the starting assumption.

### 3) Time complexity of 1D DCT algorithm

The worst case time complexity of this algorithm is clearly O(n) because maximum $(n-1)$ sequential evolutions are required in the pre-processing phase to get an intermediate configuration after which another evolution is required in the decision phase.

### 4) Experimental result

```
Input : 1 0 1 1 0 1 0 1
Preprocessing phase :
for i = 1 : 1  1  0  1  1  0  1  0
    i = 2 : 1  1  1  0  1  1  0  0
    i = 3 : 1  1  1  1  0  1  0  0
    i = 4 : 1  1  1  1  1  0  0  0
    i = 5 : 1  1  1  1  1  0  0  0
    i = 6 : 1  1  1  1  1  0  0  0
    i = 7 : 1  1  1  1  1  0  0  0
Decision phase :
(middle) and (middle+1) cell is "11"
So, Apply Rule 255 to the configuration
    i = 8 : 1  1  1  1  1  1  1  1
Output : input image is dense in 1's
```

Figure 14. Shows the pre-processing phase and the decision phase of 1D DCT algorithm for a random input string of length 8.

Fig-14 demonstrate an output configurations obtained by the 1D DCT algorithm by considering an initial 1D CA of size 8.

The following figures demonstrate the outputs obtained by the 1D DCT algorithm by considering two different initial 1D CA of size 8. Only the intermediate output is shown based on which a decision can be made.

*Input*                        *Output*
1 0 1 0 0 1 0 1 $\xrightarrow{\text{Applying 1D DCT Algorithm}}$ 1 1 1 1 0 0 0 0
                                                           *0's and 1's has equal density*

Figure 15. Shows the output of the pre-processing phase of 1D DCT algorithm. As the middle two bits are "10" so by step 31 and 32 the decision is equal density.

*Input*                        *Output*
1 0 0 0 0 1 0 1 $\xrightarrow{\text{Applying 1D DCT Algorithm}}$ 1 1 1 0 0 0 0 0
                                                         *input image is dense in 0's*

Figure 16. Shows the output of the pre-processing phase of 1D DCT algorithm. As the middle two bits are "00" so by step 28 and 29, Rule 0 is applied and the algorithm will print the CA is dense in 0's

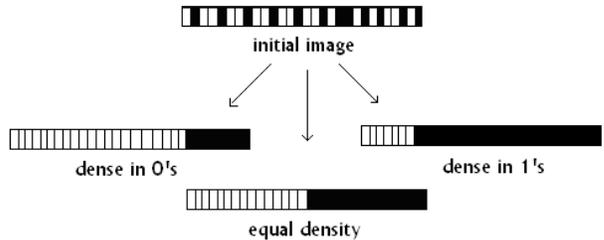

Figure 17. Possible intermediate outputs for taking decision in 1D DCT algorithm

Consider another intermediate configuration as shown in fig-17 which is the rotational symmetry of fig-9 where 1's if any in the initial CA configuration are translated from left to right instead of right to left.

Like fig-9 and table III, for fig-17, another list of rules can be found out which can also be used in the preprocessing phase of the proposed 1D DCT algorithm. These rules can be similarly obtained as before and the complete list is given in table IV. Here 1D CA Rule 184 can be obtained for the cells $x_2$ to $x_{n-1}$ as opposed to that of the CA Rule 226 earlier used for fig-9.

$XYZ$ :         111  110  101  100  011  010  001  000
Requirement :  111  101  011  010  011  001  001  000
$Rule 184$ :     1    0    1    1    1    0    0    0

Figure 18. Shows the CA rule 184 obtained by considering Translation as well as Density Preservation and is the rotational symmetry of rule 226.

Fig-19 and fig-20 shows 16 CA rules obtained by fixing the left neighbor of $x_1$ is 0 or 1 and the complete list of these rules in Wolfram's naming scheme is given in column (3) of table IV.

*Rules if left neighbour of* $x_1 = 0$
$0x_1x_2$ :      111  110  101  100  011  010  001  000
Requirement :   –    –    –    –    011  001  001  000
$16 Rules$ :    0/1  0/1  0/1  0/1   1    0    0    0

Figure 19. Shows the CA rules those can be used for the boundary cell $x_1$ if its left neighbour value is 0.

*Rules if left neighbour of* $x_1 = 1$
$1x_1x_2$ :      111  110  101  100  011  010  001  000
Requirement :  111  101  101  100   –    –    –    –
$16 Rules$ :    1    0    0    0   0/1  0/1  0/1  0/1

Figure 20. Shows the CA rules those can be used for the boundary cell $x_1$ if its left neighbour value is 1.

Similarly the rules applicable for the boundary cell $x_n$ can be found out by considering its right neighbor value either 0 or 1 as shown in fig 21 and the complete list of these rules are given in column (5) and (6) of table IV.

*Rules if right neighbour of* $x_n = 0$
$x_{n-1}x_n0$ :   111  110  101  100  011  010  001  000
Requirement :   –   110   –   010   –   010   –   000
$16 Rules$ :    0/1   1   0/1   1   0/1   1   0/1   0
(a)

*Rules if right neighbour of* $x_n = 1$
$x_{n-1}x_n1$ :   111  110  101  100  011  010  001  000
Requirement :  111   –   011   –   011   –   001   –
$16 Rules$ :     1   0/1   1   0/1   0   0/1   0   0/1
(b)

Figure 21. Shows the CA rules used for boundary cell $x_n$. (a) gives the rules if the right neighbor of $x_n = 0$ and (b) computes the rules if it is 1.

TABLE IV.     RULES USED TO SOLVE 1D DCT PROBLEM

| (1) Cells | (2) If the left neighbour of $x_1 = 0$ then apply any rule to $x_1$ | (3) If the left neighbour of $x_1 = 1$ then apply any rule to $x_1$ | (4) Rules for all other cells $x_2$ to $x_{n-1}$ | (5) If the right neighbour of $x_n = 0$ then apply any rule to $x_n$ | (6) If the right neighbour of $x_n = 1$ then apply any rule to $x_n$ |
|---|---|---|---|---|---|
| Rules | 8, 24, 40, 56, 72, 88, 104, 120, 136, 152, 168, 184, 200, 216, 232, 248 | 128, 129, 130, 131, 132, 133, 134, 135, 136, 137, 138, 139, 140, 141, 142, 143 | 184 | 84, 86, 92, 94, 116, 118, 124, 126, 212, 214, 220, 222, 244, 246, 252, 254 | 160, 161, 164, 165, 176, 177, 180, 181, 288, 289, 292, 293, 304, 305, 308, 309 |

*5) Advantage of PCA over conventional CA*

In this section the advantages of PCA over conventional CA are discussed. While solving any CA based problem with the help of a software program (or PCA) sometimes one can reduce the computing time, which cannot be avoided using standard CA. As an example, let us consider the truth table of Rule 226 as shown in fig-22.

$XYZ$ :         111  110  101  100  011  010  001  000
$Rule 226$ :     1    1    1    0    0    0    1    0
Operation :     R    R   RW    R   RW   RW   RW    R

Figure 22. Shows the CA rule 226 with the positions where only READ and where both READ and WRITE operations are required

One can observe that out of eight possible 3-neighbourhood situations four such situations viz 111, 110, 100 and 000 does not need the state change for the cell under consideration. In a sense as if for these kind of neighborhood situation only READ (R) (reading the values from neighboring cells) operation is required but no WRITE (W) operation. But for the rest four 101, 011, 010 and 001 which needs the state change both READ and WRITE (RW) operations are required. So instead of directly applying Rule 226 by looking its truth table that involves both READ and WRITE operations at every cell, if in the control unit of PCA it can be handled with the help of only READ (R) for some neighborhood situations then for many cells, the state change is not required. Therefore, using PCA one can save many WRITE operations unnecessarily involved at each cell in the entire CA evolution. To implement this idea the step 13 of the 1D DCT algorithm can be remodeled with the help of the following program fragments.

*Apply Rule 226 to the cell $x_i$*
1. {
2. *READ the neighborhood information* $(x_{i-1}, x_i, x_{i+1})$
3. *if* (*the corresponding output of* $(x_{i-1}, x_i, x_{i+1})$ *is* "*RW*") *then*
4. *WRITE the complement of* $x_i$
5. }

Further in some other situation both READ as well as WRITE operations can be avoided. As an example when few 1's are present in the right side region of the initial CA configuration as shown in fig-23, it can be observed that when rules are applied in the preprocessing phase, not all cells are required to evolve in each evolution process. Thus in the control unit of PCA programmatically one can reduce the CA size in each successive evolution and can save a lot of READ and WRITE operations.

*Initial*:  0 0 0 0 0 0 0 0 0 0 1 1
*for i* =1: 0 0 0 0 0 0 0 0 0 1 0 1
*i* =2: 0 0 0 0 0 0 0 0 1 0 1 0
*i* =3: 0 0 0 0 0 0 0 1 0 1 0 0
*i* =4: 0 0 0 0 0 0 1 0 1 0 0 0
*i* =5: 0 0 0 0 0 1 0 1 0 0 0 0
*i* =6: 0 0 0 0 1 0 1 0 0 0 0 0
*i* =7: 0 0 0 1 0 1 0 0 0 0 0 0
*i* =8: 0 0 1 0 1 0 0 0 0 0 0 0
*i* =9: 0 1 0 1 0 0 0 0 0 0 0 0
*i* =10: 1 0 1 0 0 0 0 0 0 0 0 0
*i* =11: 1 1 0 0 0 0 0 0 0 0 0 0

Figure 23. Shows all the iterations in the preprocessing phase of 1D DCT algorithm. Here all the CA evolutions are shown starting from an initial CA configuration of size 12. As in the last iteration, two cells in the middle are 00 so Rule 0 is applied in the decision phase to reach at a configuration of all 0's.

It may be noted that it is often difficult to implement conventional CA, which can solve simple computing tasks, such as DCT. Here the authors propose a solution to the DCT by using a variant of CA, called Programmable CA (PCA), in which the automaton is allowed to change the updating rule both in space and time, i.e. different components can use different updating rules and they are allowed to change their updating rule during the computation. It is natural to get the criticism about the approach adopted in the said algorithm. The main difficulty of this PCA approach is that the solution must be computed by simple interacting devices which can observe the state of their neighbors. These devices are further required to organize themselves to find a solution with the help of some external entity, which has the knowledge like whether the length of the CA is odd or even. If odd, the status of the middle cell and if even then the status of the middle two cells, will be required. Further we have to know which are the inner cells or boundary cells.

The pseudo-code proposed is very much relies on this information and given all such information available the self-organizing behavior of CA can gradually reach to the DCT solution. Thus it is quite possible that some smart implementation of PCA would overcome this problem by suitably providing this information and a dedicated hardware device or Cellular Automata Machine (CAM) (Toffoli and Margolis, 1987) can be programmed such that by its control unit many READ and WRITE operations of CA rules can be avoided and a DCT solution can be arrived at. Author's earlier papers (Sahoo and Choudhury, 2008; Choudhury, Sahoo, Chakraborty, 2008) can give certain directions to design such a Cellular Automata Machine (CAM) for the PCA.

### 6) General 1D DCT Solution

The 1D DCT algorithm discussed above can be slightly modified so that the general 1D DCT problem for arbitrary critical density $\rho_c$ can be solved. After the preprocessing phase is over in the proposed 1D DCT algorithm, for an odd length CA *n* by looking the $(n/c)^{th}$ position of the vector $(x_1, x_2, x_3, ..., x_{n-1}, x_n)$ either 0 or 1 one can take the decision of their densities. For example if c=1/3 and *n* is odd, then the finding value of $x_{n/3}$=0 or 1 one can take the decision as dense in 0's or dense in 1's respectively. Similarly, for even length CA two cell values at $(n/c)^{th}$ and $(n/(c+1))^{th}$ position of the vector $(x_1, x_2, x_3, ..., x_{n-1}, x_n)$ decides the solution. Thus the value c acts as a scale of measurement for taking the decision to solve the general DCT problem.

One such example for the general density classification problem of density $\rho_c$ where c=1/3 is the electrolysis process of $H_2O$ or water. During the process of electrolysis $(2/3)^{rd}$ Hydrogen molecules and $(1/3)^{rd}$ Oxygen molecules are stored in anode and cathode respectively. Thus the preprocessing phase embedded in the definition is a natural way to solve a general 1D DCT problem. That is two corners of the initial 1D array can be thought of as two electrodes, cathode and anode and the charge particles -ve and +ve charges (say 1's and 0's respectively) are attracted by their corresponding electrodes so that all the +ve ions are stored in the anode and the –ve charges are stored in the cathode. Similar analogy can be made for magnetism considering two corners as two poles North Pole and South Pole of a magnet. Above all the main objective of solving the DCT problem is to better understand the limitations of CA.

## VI. SOLUTION OF 2D DCT USING PROGRAMMABLE CELLULAR AUTOMATA (PCA)

### A. Algorithm to solve 2D Density Classification problem

In this section an algorithm is proposed using PCA rules to solve 2D DCT problem. Here 9 fundamental rules given in

fig-24 are used to achieve both translations of images as well as the density preservation throughout the CA evolution. This algorithm is also a two-phase procedure like 1D DCT algorithm given in section V. The technique used in the preprocessing phase is a slight modification of Sweepers algorithm reported in our earlier paper (Choudhury et al., 2008) and the name is so because the algorithm is very similar to the way a sweeper sweeps haphazardly and puts it in one corner of a room. The cells those are in the boundary region towards the direction of translation are taken as the points of destination and on repeated application of fundamental 2D CA rules, all 1's comes closer to these points. For example in the pre-processing phase to achieve a configuration like fig-24 in a $(n \times n)$ matrix, the boundary cells lies in the cell position $(i,1)$ or $(1, j)$ for all $1 \leq i, j \leq n$ can be considered as the points of destination.

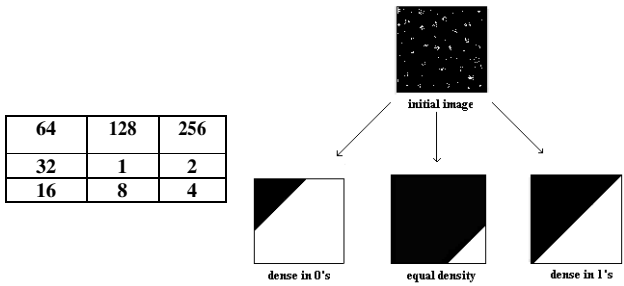

Figure 24. Shows an intermediate configuration for 2D DCT solution using Programmable CA rules 1, 2, 4, 8, 16, 32, 64, 128 and 256

Let the variable $K$ be a fundamental 2D CA rule such as 2, 4, 8, 16 and $K^T$ be its corresponding transpose rule like 32, 64, 128, 256 respectively. From table II of section IV, Rule 1 can be used as density preservation where as other 8 rules can be used for translation of 1's and 0's in different directions but the directions of translation for Rule $K$ and $K^T$ are opposite to each other. Notice that Rules $(1, K, K^T)$ in fig-24 are always lies in 1D. So, the combined effect of these 3 fundamental rules $(1, K, K^T)$ can be iteratively applied to any 1D configuration to get an image like fig-9. As any 2D matrix can be divided into several 1D arrays either in rows, columns or in diagonals so the pre-processing phase of 2D DCT algorithm shown in fig-24 can be visualized as several pre-processing figures like fig-9 used for 1D DCT. The general code for these 3 fundamental rules $(1, K, K^T)$ applied to a 2D matrix of size $(n \times n)$ is shown next followed by an example shown in fig-25 where the rules (1, 4, 64) and (1, 16, 256) are applied to an arbitrary 2D binary image of size $(6 \times 6)$.

/*Input-A 2D *binary* image of size $(n \times n)$. Just like 1D, 3-neighborhood CA, consider the variable Y be the Cell under consideration whose left and right neighbors are X and Z respectively.*/

Apply Rules $(1, K, K^T)$
1. {
// Boundary cells in the translation direction of Rule K
2. *if* Y=1 then Apply Rule 1 and *if* Y=0 Apply Rule K
//All Other Cells
3. *if* (the 3-Neighborhood in the translation direction of Rule K is 110)
4. *then* Apply Rule 1        /*Density preservation*/
5. *else if* (the 3-Neighborhood in the translation direction of Rule K is 011)
6. *then* Apply Rule $K^T$  /* 0's moves in the translation direction of Rule $K^T$*/
7. *else* Apply Rule K        /* 1's moves in the translation direction of Rule K*/
}

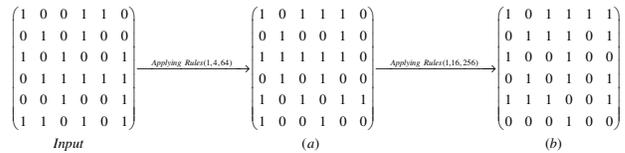

Figure 25. (a) shows the function Apply Rules $(1, 4, 64)$ is applied to an arbitrary 2D image of size $(6 \times 6)$ and as a result 1's are translated towards the traslation direction of Rule 4 i.e the direction shown in fig-24. similarly (b) shows the final output when the function Apply Rules $(1, 16, 256)$ and Apply Rules $(1, 4, 64)$ are applied to the initial image one after another.

The algorithm using the above function for solving 2D DCT problem corresponding to the intermediate configuration shown in fig-24 is given next.

### 1) Algorithm to solve 2D DCT

/* Preprocessing Phase : Reaching to an intermediate configuration */

1. A 2D *binary* image of size $(n \times n)$ is taken as the input and just like 1D, 3-neighborhood CA consider the variable Y be the Cell under consideration whose left and right neighbors are X and Z respectively.
2. *for* i ← 1 to n
3. {
4. Apply Rules $(1, 4, 64)$
5. Apply Rules $(1, 16, 256)$
6. }
7. *for* i ← 1 to n
8. {
9. Apply Rules $(1, 16, 256)$
10. Apply Rules $(1, 8, 128)$
11. Apply Rules $(1, 2, 32)$
12. }
13. *for* i ← 1 to n
14. {
15. Apply Rules $(1, 16, 256)$ /*Rules applied to the decision line only*/
16. }

/* Decision phase : For taking a decision of density */

17. *if n is odd*
18. {
19.    *if* the bit position $((n+1)/2,(n+1)/2) = 1$ *then*
20.      Apply Rule $(2^{512}-1)$ to all the cells
21.      *print* "the image is dense in 1's".
22.    *else if* the bit position $((n+1)/2, (n+1)/2) = 0$ *then*
23.      Apply Rule 0 to all the cells
24.      *print* "the image is dense in 0's"
25. }
26. *else if n is even*
27. {
28.    *if* the bit position $((n/2)+1, n/2) = 0$ and $(n/2,(n/2)+1) = 0$ *then*
29.      Apply Rule 0 to all the cells
30.      *print* "the image is dense in 0's".
31.    *else if* the bit position $((n/2)+1, n/2) = 1$ and $(n/2,(n/2)+1) = 0$ *then*
32.      *print* "equal density in 0's and 1's".
33.    *else if* the bit position $((n/2)+1, n/2) = 1$ and $(n/2,(n/2)+1) = 1$ *then*
34.      Apply Rule $(2^{512}-1)$ to all the cells
35.      *print* "the image is dense in 1's".
36. }

### 2) Explanation of the algorithm

The proposed algorithm for both density preservation and translation of 1's towards one half of the region uses the following logic.

If the boundary cells towards the translation direction are 0 then only the CA rule is applied. For example in step 4 of the preprocessing phase when the function $Apply\ Rules\ (1, K, K^T)$ is called for Rules 1, 4 and 64 then Rule 4 is applied to the boundary cells lies in the cell position $(i,1)$ or $(1, j)$ for all $1 \le i, j \le n$ where the cell values are 0. In addition to this, Rule 4 is also applied to all other cells except if the cell, its left neighbor and right neighbor values are "110" and "011". While in the first case "110" Rule 1 is applied whereas, in the second "011" the transpose rule 4 which is Rule 64 is applied. The situation in 2D nine-neighborhood for Rule 1, Rule 4 and Rule 64 is shown in fig-26 by treating it as a 1D 3-neighborhood CA having neighborhood variables X, Y and Z as usual.

|   |   |   |
|---|---|---|
| X | 0/1 | 0/1 |
| 0/1 | Y | 0/1 |
| 0/1 | 0/1 | Z |

(a)

|   |   |   |
|---|---|---|
| 1 | 0/1 | 0/1 |
| 0/1 | 1 | 0/1 |
| 0/1 | 0/1 | 0 |

(b)

|   |   |   |
|---|---|---|
| 0 | 0/1 | 0/1 |
| 0/1 | 1 | 0/1 |
| 0/1 | 0/1 | 1 |

(c)

Figure 26. (a) Shows 2D 9- neighborhood CA and Rule 4 is applied to all possible 9 neighborhoods except the string XYZ is 110 or 011. In (b) Rule 1 is applied if XYZ is 110 and in (c) Rule 64 is applied if XYZ is 011.

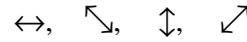

Figure 27. 2D 9- neighborhoods can be considered as 1D 3-neighborhood CA with different orientations

In the pre-processing phase, 2D nine-neighborhoods can be considered as four 1D 3-neighborhoods in different orientations as shown in fig-27. Thus the rules (32, 1, 2), (64, 1, 4), (128, 1, 8) and (16, 1, 256) are iteratively applied to a 2D binary image as shown from step 2 to step 12 of the algorithm to get an intermediate configuration like fig-24. In steps 13 to 16 the function $Apply\ Rules\ (1,16,256)$ is called and the rules are applied only to the cells lies in the diagonal line (decision line) after which a decision can be made in the decision phase like 1D solution. Different intermediate outputs of this algorithm given in fig-28 self explains how this sweeper's algorithm works.

It can be noted that the function $Apply\ Rules\ (1, K, K^T)$ is written in whatever form is same as the 1D CA rules given in table III and in table IV. Therefore, in the pre-processing phase of 2D DCT algorithm Rule 226 can be applied repeatedly to each cell by considering different 3 neighborhoods like (16, 1, 256), (64, 1, 4), (128, 1, 8) and (16, 1, 256) one after another except the boundary cells at first row and first column positions. To the boundary cells any one of the 16 possible rules is applied as given in table III and table IV. Finally, a trivial rule: Rule 0 or Rule $(2^{512}-1)$ is applied to get the desired solution of 2D DCT.

### 3) Time complexity of 2D DCT algorithm

The time complexity of this CA algorithm is O(n) so computationally better than the sequential algorithm for counting number of 1's and taking the decision which generally takes $O(n^2)$ time.

### 4) Experimental Results

The following figures demonstrate the three possibilities for 2D DCT by considering different initial configurations of size (6 x 6). Fig-28 shows the output of different intermediate steps (Steps 1 to 16) of the algorithm after which a decision is made in the decision phase.

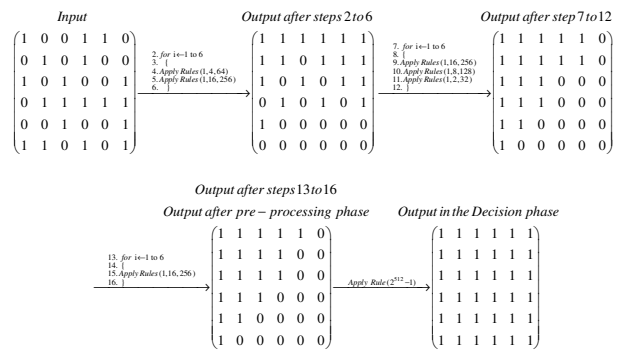

Figure 28. In the pre-processing phase output, the bits at cell position (4, 3)=1 and (3, 4)=1. So by step 34 of 2D DCT algorithm the initial image is dense in 1's

Fig-29 and fig-30 shows the output of both pre-processing phase and the decision phase starting from a random initial configuration.

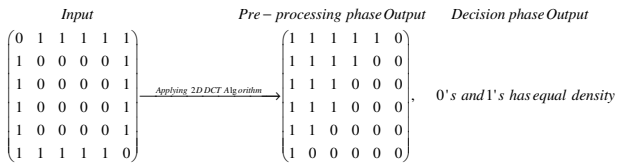

Figure 29. In the pre-processing phase output, the bits at cell position (4, 3)=1 and (3, 4)=0. So by step 32 of 2D DCT algorithm the density of 0's and 1's are same.

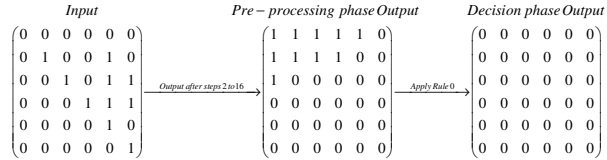

Figure 30. In the pre-processing phase output, the bits at cell position (4, 3)=0 and (3, 4)=0. So by step 30 of 2D DCT algorithm the initial image is dense in 0's

Just like 1D, other 2D images can be used as the output of the pre-processing phase as shown in fig-31 and fig-32; after which a decision of density can be made in the decision phase. Similar algorithms like the previous 2D DCT algorithm can also be designed for these figures. It may be noted that for a square image of size $(n \times n)$ all these figures can able to solve the 2D DCT problem. But for an arbitrary rectangular image of size $(m \times n)$ only the figures given in fig-32 can be used as an output of the pre-processing phase because for a rectangular image a diagonal region for taking the decision of classification cannot be obtained.

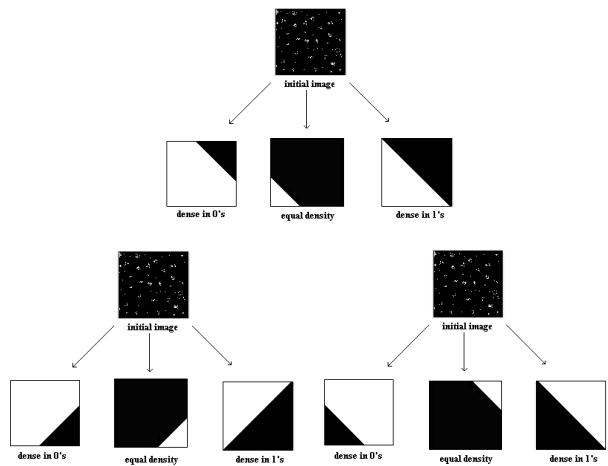

Figure 31. Shows other intermediate configurations can be used in the pre-processing phase of the 2D DCT solution

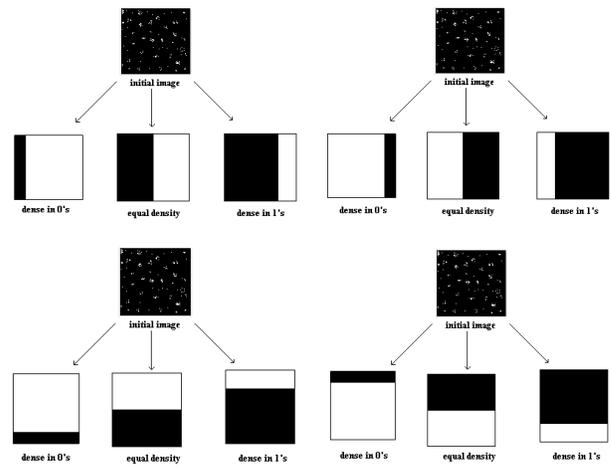

Figure 32. Shows other intermediate configurations can be used in the pre-processing phase of the 2D DCT solution

### 5) *Variety of 2D density classification problem*

In this section some variants of 2D DCT problem is defined. Their applications and the solution using the similar kind of logic using PCA are also discussed.

The conventional 2D DCT problem is defined on a rectangular or a square shape initial CA configuration. Instead of this one can define the problem on various symmetric images like triangular, circular, parallelogram etc. The PCA and the two-phase method used previously to solve the conventional 2D DCT problem can be similarly used to solve this type of problems. Also it is possible to define the 2D DCT on an asymmetric initial CA configuration. But for its solution instead of linear separation, the algorithm can be modified by dividing the region by a curve into two regions of equal area each, and then the fundamental rules are selected to sweep all the 1's to bring it to a single region after which a decision of classification can be made.

Up to this the DCT is defined over a continuous image. Now another variant of DCT is to compute the density in a discontinuous image. The application of this problem might be, computing the density of literate and illiterate people of a place, computing the total density of male and female of a country on knowing individual density of each state and so on. As exact number of literate and illiterate people in different regions are generally unknown a priory so it may be a difficult task to solve this type of problem using standard CA. But, the pre-processing phase used in this paper can act as a scale of measurement to find the exact density in percentage and can able to solve this type of problem. In our future research we will be dealing these new kinds of DCT in detail.

## VII. CONCLUSION AND FUTURE RESEARCH DIRECTIONS

This paper highlights some existing solution of both 1D and 2D DCT reported previously by various authors using conventional CA and given its solution using a variant called Programmable Cellular Automata (PCA). The approach adopted here is a two phase procedure namely pre-processing phase and the decision phase. In the pre-processing phase a set of hybrid CA rules are applied iteratively to an initial CA to reach at a configuration after which the density decision is made in the decision phase. The proposed DCT algorithm is very much compatible in order to handle different boundary conditions such as fixed, null, periodic, intermediate, adiabatic, reflexive etc. The advantage of PCA over traditional CA rules is discussed and the general DCT for arbitrary 2D image of various shapes and sizes, its applicability and solution using PCA are first time introduced. Regarding the future effort, it is the strong belief of the authors that the proposed approach has several other analogies useful for many physical, chemical and mathematical simulations. Some of these are as follows:

1. A connection can be established between Iterated Function System (IFS) and the logic used in the pre-processing phase. Here two corner points can be the point of attraction.

2. Also an effort can be made to establish a connection between Multiple Attractor Cellular Automata (MACA) with the proposed DCT algorithm to solve other pattern classification problem. Here also two corner points can be considered as two basins of attractors and all 1's form a cluster in one basin where as all 0's can form a cluster of another basin.

3. The above algorithm linearly separates the clusters may be useful to solve other pattern classification problems except DCT.

4. It can be thought of as a 2D filter that may be useful for separating the noise from the actual image.

Exploring all these applications using 1D and 2D DCT is our immediate future efforts.